\RequirePackage{lineno}
\documentclass[aps,prd,twocolumn,superscriptaddress,showpacs,amsmath,amssymb]{revtex4}

\usepackage{comment}
\usepackage{multirow}
\usepackage{dcolumn}
\usepackage{bm}
\usepackage{lineno}
\usepackage{setspace}
\usepackage{graphicx}
\includecomment{printrobustnotes}
\includecomment{printallnotes}
\usepackage{xspace}

\usepackage{dcolumn}
\usepackage{bm}

\newenvironment{cfigure1c}[1][tbp]{\begin{figure*}[#1]\centering}{\end{figure*}}

\newcommand{\note}[1]{\xspace}
\newcommand{\robustnote}[1]{\xspace}
\begin{printrobustnotes}
\renewcommand{\robustnote}[1]{\textbf{\textit{#1}}\xspace}
\end{printrobustnotes}
\begin{printallnotes}
\renewcommand{\note}[1]{\textbf{\textit{#1}}\xspace}
\renewcommand{\robustnote}[1]{\textbf{\textit{#1}}\xspace}
\end{printallnotes}
\newcommand{\ppbar}  {\ensuremath{p\bar{p}}\xspace}
\newcommand{\ttbar}  {\ensuremath{t\bar{t}}\xspace}

\newcommand{\Ht}     {\ensuremath{H_{\mathrm{T}}}\xspace}
\newcommand{\et}     {\ensuremath{E_{\mathrm{T}}}\xspace}
\newcommand{\etl}     {\ensuremath{E_{\mathrm{T}}^{\mathrm{lepton}}}\xspace}

\newcommand{\etj}     {\ensuremath{E_{\mathrm{T}}^{\mathrm{jet}}}\xspace}

\newcommand{\pt}     {\ensuremath{p_{\mathrm{T}}}\xspace}
\newcommand{\ptt}     {\ensuremath{p_{\mathrm{T}}^{\mathrm{total}}}\xspace}
\newcommand{\ptl}     {\ensuremath{p_{\mathrm{T}}^{\mathrm{lepton}}}\xspace}

\newcommand{\met}    {\ensuremath{\text{\raisebox{.3ex}{$\not$}}\et}\xspace}

\newcommand{\mtop}    {\ensuremath{{M}_{\text{top}}}\xspace}

\newcommand{\mtr}     {\ensuremath{m_{\text{t}}^{\text{reco}}}\xspace}
\newcommand{\mttr}     {\ensuremath{m_{\text{t}}^{\text{reco(2)}}}\xspace}

\newcommand{\mjj}    {\ensuremath{m_{\mathrm{jj}}}\xspace}

\newcommand{\djes}   {\ensuremath{\Delta_{\mathrm{JES}}}\xspace}

\newcommand{\genunit}[2]{\ensuremath{#1~\mathrm{#2}}\xspace}

\newcommand{\gev}[1]    {\genunit{#1}{GeV}}
\newcommand{\pb}[1]     {\genunit{#1}{pb}}

\newcommand{\gevc}[1]   {\ensuremath{#1~\mathrm{GeV}/c}}
\newcommand{\gevcc}[1]  {\ensuremath{#1~\mathrm{GeV}/c^{2}}}

\newcommand{\invfb}[1]  {\ensuremath{#1~\mathrm{fb}^{-1}}}

\newcommand{\gevccnoarg}{\ensuremath{\mathrm{GeV}/c^{2}}\xspace}

\newcommand{\measErr}[2]{\ensuremath{#1 \pm #2}\xspace}

\newcommand{\measStatSyst}[3]{\ensuremath{#1 \pm #2~(\textrm{stat}) \pm #3~(\textrm{syst})}\xspace}

\begin{document}

\title{Precision Top-Quark Mass Measurement at CDF}

\affiliation{Institute of Physics, Academia Sinica, Taipei, Taiwan 11529, Republic of China}
\affiliation{Argonne National Laboratory, Argonne, Illinois 60439, USA}
\affiliation{University of Athens, 157 71 Athens, Greece}
\affiliation{Institut de Fisica d'Altes Energies, ICREA, Universitat Autonoma de Barcelona, E-08193, Bellaterra (Barcelona), Spain}
\affiliation{Baylor University, Waco, Texas 76798, USA}
\affiliation{Istituto Nazionale di Fisica Nucleare Bologna, $^{ee}$University of Bologna, I-40127 Bologna, Italy}
\affiliation{University of California, Davis, Davis, California 95616, USA}
\affiliation{University of California, Los Angeles, Los Angeles, California 90024, USA}
\affiliation{Instituto de Fisica de Cantabria, CSIC-University of Cantabria, 39005 Santander, Spain}
\affiliation{Carnegie Mellon University, Pittsburgh, Pennsylvania 15213, USA}
\affiliation{Enrico Fermi Institute, University of Chicago, Chicago, Illinois 60637, USA}
\affiliation{Comenius University, 842 48 Bratislava, Slovakia; Institute of Experimental Physics, 040 01 Kosice, Slovakia}
\affiliation{Joint Institute for Nuclear Research, RU-141980 Dubna, Russia}
\affiliation{Duke University, Durham, North Carolina 27708, USA}
\affiliation{Fermi National Accelerator Laboratory, Batavia, Illinois 60510, USA}
\affiliation{University of Florida, Gainesville, Florida 32611, USA}
\affiliation{Laboratori Nazionali di Frascati, Istituto Nazionale di Fisica Nucleare, I-00044 Frascati, Italy}
\affiliation{University of Geneva, CH-1211 Geneva 4, Switzerland}
\affiliation{Glasgow University, Glasgow G12 8QQ, United Kingdom}
\affiliation{Harvard University, Cambridge, Massachusetts 02138, USA}
\affiliation{Division of High Energy Physics, Department of Physics, University of Helsinki and Helsinki Institute of Physics, FIN-00014, Helsinki, Finland}
\affiliation{University of Illinois, Urbana, Illinois 61801, USA}
\affiliation{The Johns Hopkins University, Baltimore, Maryland 21218, USA}
\affiliation{Institut f\"{u}r Experimentelle Kernphysik, Karlsruhe Institute of Technology, D-76131 Karlsruhe, Germany}
\affiliation{Center for High Energy Physics: Kyungpook National University, Daegu 702-701, Korea; Seoul National University, Seoul 151-742, Korea; Sungkyunkwan University, Suwon 440-746, Korea; Korea Institute of Science and Technology Information, Daejeon 305-806, Korea; Chonnam National University, Gwangju 500-757, Korea; Chonbuk National University, Jeonju 561-756, Korea; Ewha Womans University, Seoul, 120-750, Korea}
\affiliation{Ernest Orlando Lawrence Berkeley National Laboratory, Berkeley, California 94720, USA}
\affiliation{University of Liverpool, Liverpool L69 7ZE, United Kingdom}
\affiliation{University College London, London WC1E 6BT, United Kingdom}
\affiliation{Centro de Investigaciones Energeticas Medioambientales y Tecnologicas, E-28040 Madrid, Spain}
\affiliation{Massachusetts Institute of Technology, Cambridge, Massachusetts 02139, USA}
\affiliation{Institute of Particle Physics: McGill University, Montr\'{e}al, Qu\'{e}bec, Canada H3A~2T8; Simon Fraser University, Burnaby, British Columbia, Canada V5A~1S6; University of Toronto, Toronto, Ontario, Canada M5S~1A7; and TRIUMF, Vancouver, British Columbia, Canada V6T~2A3}
\affiliation{University of Michigan, Ann Arbor, Michigan 48109, USA}
\affiliation{Michigan State University, East Lansing, Michigan 48824, USA}
\affiliation{Institution for Theoretical and Experimental Physics, ITEP, Moscow 117259, Russia}
\affiliation{University of New Mexico, Albuquerque, New Mexico 87131, USA}
\affiliation{The Ohio State University, Columbus, Ohio 43210, USA}
\affiliation{Okayama University, Okayama 700-8530, Japan}
\affiliation{Osaka City University, Osaka 588, Japan}
\affiliation{University of Oxford, Oxford OX1 3RH, United Kingdom}
\affiliation{Istituto Nazionale di Fisica Nucleare, Sezione di Padova-Trento, $^{ff}$University of Padova, I-35131 Padova, Italy}
\affiliation{University of Pennsylvania, Philadelphia, Pennsylvania 19104, USA}
\affiliation{Istituto Nazionale di Fisica Nucleare Pisa, $^{gg}$University of Pisa, $^{hh}$University of Siena and $^{ii}$Scuola Normale Superiore, I-56127 Pisa, Italy}
\affiliation{University of Pittsburgh, Pittsburgh, Pennsylvania 15260, USA}
\affiliation{Purdue University, West Lafayette, Indiana 47907, USA}
\affiliation{University of Rochester, Rochester, New York 14627, USA}
\affiliation{The Rockefeller University, New York, New York 10065, USA}
\affiliation{Istituto Nazionale di Fisica Nucleare, Sezione di Roma 1, $^{jj}$Sapienza Universit\`{a} di Roma, I-00185 Roma, Italy}
\affiliation{Rutgers University, Piscataway, New Jersey 08855, USA}
\affiliation{Texas A\&M University, College Station, Texas 77843, USA}
\affiliation{Istituto Nazionale di Fisica Nucleare Trieste/Udine, I-34100 Trieste, $^{kk}$University of Udine, I-33100 Udine, Italy}
\affiliation{University of Tsukuba, Tsukuba, Ibaraki 305, Japan}
\affiliation{Tufts University, Medford, Massachusetts 02155, USA}
\affiliation{University of Virginia, Charlottesville, Virginia 22906, USA}
\affiliation{Waseda University, Tokyo 169, Japan}
\affiliation{Wayne State University, Detroit, Michigan 48201, USA}
\affiliation{University of Wisconsin, Madison, Wisconsin 53706, USA}
\affiliation{Yale University, New Haven, Connecticut 06520, USA}

\author{T.~Aaltonen}
\affiliation{Division of High Energy Physics, Department of Physics, University of Helsinki and Helsinki Institute of Physics, FIN-00014, Helsinki, Finland}
\author{B.~\'{A}lvarez~Gonz\'{a}lez$^z$}
\affiliation{Instituto de Fisica de Cantabria, CSIC-University of Cantabria, 39005 Santander, Spain}
\author{S.~Amerio}
\affiliation{Istituto Nazionale di Fisica Nucleare, Sezione di Padova-Trento, $^{ff}$University of Padova, I-35131 Padova, Italy}
\author{D.~Amidei}
\affiliation{University of Michigan, Ann Arbor, Michigan 48109, USA}
\author{A.~Anastassov$^x$}
\affiliation{Fermi National Accelerator Laboratory, Batavia, Illinois 60510, USA}
\author{A.~Annovi}
\affiliation{Laboratori Nazionali di Frascati, Istituto Nazionale di Fisica Nucleare, I-00044 Frascati, Italy}
\author{J.~Antos}
\affiliation{Comenius University, 842 48 Bratislava, Slovakia; Institute of Experimental Physics, 040 01 Kosice, Slovakia}
\author{G.~Apollinari}
\affiliation{Fermi National Accelerator Laboratory, Batavia, Illinois 60510, USA}
\author{J.A.~Appel}
\affiliation{Fermi National Accelerator Laboratory, Batavia, Illinois 60510, USA}
\author{T.~Arisawa}
\affiliation{Waseda University, Tokyo 169, Japan}
% ...
\collaboration{CDF Collaboration\footnote{With visitors from
$^a$Istituto Nazionale di Fisica Nucleare, Sezione di Cagliari, 09042 Monserrato (Cagliari), Italy,
$^b$University of CA Irvine, Irvine, CA 92697, USA,
$^c$University of CA Santa Barbara, Santa Barbara, CA 93106, USA,
$^d$University of CA Santa Cruz, Santa Cruz, CA 95064, USA,
$^e$Institute of Physics, Academy of Sciences of the Czech Republic, Czech Republic,
$^f$CERN, CH-1211 Geneva, Switzerland,
$^g$Cornell University, Ithaca, NY 14853, USA,
$^h$University of Cyprus, Nicosia CY-1678, Cyprus,
$^i$Office of Science, U.S. Department of Energy, Washington, DC 20585, USA,
$^j$University College Dublin, Dublin 4, Ireland,
$^k$ETH, 8092 Zurich, Switzerland,
$^l$University of Fukui, Fukui City, Fukui Prefecture, Japan 910-0017,
$^m$Universidad Iberoamericana, Mexico D.F., Mexico,
$^n$University of Iowa, Iowa City, IA 52242, USA,
$^o$Kinki University, Higashi-Osaka City, Japan 577-8502,
$^p$Kansas State University, Manhattan, KS 66506, USA,
$^r$University of Manchester, Manchester M13 9PL, United Kingdom,
$^s$Queen Mary, University of London, London, E1 4NS, United Kingdom,
$^t$University of Melbourne, Victoria 3010, Australia,
$^u$Muons, Inc., Batavia, IL 60510, USA,
$^v$Nagasaki Institute of Applied Science, Nagasaki, Japan,
$^w$National Research Nuclear University, Moscow, Russia,
$^x$Northwestern University, Evanston, IL 60208, USA,
$^y$University of Notre Dame, Notre Dame, IN 46556, USA,
$^z$Universidad de Oviedo, E-33007 Oviedo, Spain,
$^{aa}$CNRS-IN2P3, Paris, F-75205 France,
$^{bb}$Texas Tech University, Lubbock, TX 79609, USA,
$^{cc}$Universidad Tecnica Federico Santa Maria, 110v Valparaiso, Chile,
$^{dd}$Yarmouk University, Irbid 211-63, Jordan,
}}
\noaffiliation

\date{\today}
\begin{abstract}
	We present a precision measurement of the top-quark mass using the full sample of Tevatron $\sqrt{s}=1.96$~TeV proton-antiproton collisions collected by the CDF~II detector, corresponding to an integrated luminosity of \invfb{8.7}. Using a sample of \ttbar candidate events decaying into the lepton+jets channel, we obtain distributions of the top-quark masses and the invariant mass of two jets from the $W$ boson decays from data. We then compare these distributions to templates  derived from signal and background samples to extract the top-quark mass and the energy scale of the calorimeter jets with {\it in situ} calibration. The likelihood fit of the templates from signal and background events to the data yields the single most-precise measurement of the top-quark mass, $\mtop = \gevcc{\measStatSyst{172.85}{0.71}{0.85}}.$
\end{abstract}

\maketitle

The top quark~($t$) is by far the heaviest known elementary particle~\cite{pdg}. It contributes significantly to electroweak radiative 
corrections relating the top-quark mass~($\mtop$) and the $W$ boson mass to the mass of the Higgs boson~\cite{elfit}. 
Precision measurements of $\mtop$ provide therefore important constraints on the Higgs boson mass. 
Over the last decade, the CDF and D0 experiments have been improving the precision of the \mtop measurement~\cite{top_average}, joined recently by the experiments at the Large Hadron Collider~\cite{lhc_topmass}. 

This Letter reports the single most-precise measurement of the top-quark mass to date using the template method~\cite{tmt1,tmt2,tmt3,tmt4}. In this measurement, we reconstruct  top-quark masses in each event and compare the distribution of data with template distributions derived from model calculations to estimate \mtop. We also use the template distributions of hadronically-decaying $W$ bosons to constrain the jet energy scale, an important uncertainty in the \mtop measurement. This is an update of a previous measurement that used a subset of the present data and determined \mtop=\gevcc{172.2 \pm 1.5}~\cite{tmt4}. 
This measurement has an accuracy approximately 30\% better than an earlier result with similar template technique and 12\% better with respect to the previous best measurement~\cite{mtm}. In this measurement, we use not only larger samples but also improve jet energy calibration using an artificial neural network~\cite{nnjes} to achieve better jet energy resolution, and increase signal acceptance allowing less pure signal samples into the analysis. 
The measurement is performed with $\sqrt{s} = 1.96$~TeV Tevatron proton-antiproton collision data collected by the CDF II detector~\cite{r_cdfDetector} and corresponding to an integrated luminosity of \invfb{8.7}. This is the full Run~II CDF data set with requirements of good performances in the all detector elements. 
  
Assuming unitarity of the quark-mixing matrix~\cite{pdg}, the top quark decays almost exclusively into a $W$ boson and a $b$ quark. 
In the \ttbar events, the case where one $W$ decays leptonically into an electron~($e$) or a muon~($\mu$) plus a neutrino~($\nu$), including the cascade decay of $W \rightarrow \tau \nu$ and $\tau \rightarrow \mu \nu$ or $\tau \rightarrow e\nu$, and the other $W$
decays hadronically into a pair of jets~($\ttbar \rightarrow l\nu b\bar{b}q\bar{q}$) defines the lepton+jets decay channel. 
Lepton+jets events are selected by requiring one isolated~\cite{isolated} electron (muon) with 
$\et > \gev{20}$ ($\pt > \gevc{20}$) and pseudorapidity 
$|\eta|<1.1$~\cite{cdfco}. 
We also require large missing transverse energy~\cite{met}, $\met > \gev{20}$, and at least four jets. 
Jets are reconstructed with a cone algorithm~\cite{jetclu} with radius 
$R = \sqrt{(\Delta \eta)^2+(\Delta \phi)^2} = 0.4$. 
In addition to the standard jet energy corrections~\cite{jes}, we train a neural network including additional information to the calorimeter one, such as jet momentum from the tracker as described in Ref.~\cite{nnjes}. We have performed the training separately for $b$ quarks and light flavor quarks from $W$ boson decay. The additional information on the jets improves the resolution of the reconstructed top-quark masses and hadronically decaying $W$ boson mass. This allows approximately 13\% better statistical precision compared with standard correction. 
Jets originating from $b$ quarks are identified~(tagged) using a secondary-vertex tagging algorithm~\cite{secvtx}. 
We divide the sample of candidate lepton+jets events into sub-samples based on the number of identified $b$ jets, zero $b$-tagged jet~(0-tag), one $b$-tagged jet~(1-tag), and two or more $b$-tagged jets~(2-tag). 
In the 0-tag events, we require exactly four tight jets (transverse energy $\et > \gev{20}$  and $|\eta|<2.0$). In the 1-tag and 2-tag events, three tight jets and at least one loose jet ($\et > \gev{12}$ and $|\eta|<2.4$) are required. 
We divide the 1-tag and 2-tag samples into sub-samples based on the number of tight jets and call the ``tight'' sub-sample the one requiring exactly four tight jets and ``loose'' the one consisting of the remaining events passing selection. The measurement uses five sub-samples~(0-tag, 1-tagL, 1-tagT, 2-tagL, and 2-tagT, where L and T represent loose and tight selection, respectively).  The sub-samples of 0-tag and 1-tagL are newly added in this measurement. The introduction of these higher-background sub-samples offers a gain of approximately 12\% in the statistical precision. 
We apply an additional requirement on the scalar sum of transverse energies 
in the event, $\Ht = \etl + \met + \sum_{\mathrm{jets}}\etj$~\cite{Htcut}, to further reject backgrounds, where $\etl$ is electron or muon transverse energy or momentum, respectively, and $\etj$ is the transverse energy of the jet. 

The primary background sources are $W$+jets and QCD multijet production. 
We also consider small contributions from $Z$+jets, diboson, and single-top-quark production. 
To estimate the contribution of each process, we use a combination of data- and Monte Carlo~(MC)-based 
techniques described in Ref.~\cite{xsection}. For the $Z$+jets, diboson, single top, and $\ttbar$ events
we normalize simulated events using their theoretical cross sections~\cite{smcross1,smcross2,thcross}. 
We use the data-driven techniques described in Ref.~\cite{qcdback} to estimate QCD multijet background. The shape of the $W$+jets background is obtained from simulation, while the number of $W$+jets 
events is determined from the data by subtracting the other backgrounds and $\ttbar$ event contributions. 

For each event, three observables are used, two reconstructed top-quark masses~(\mtr and \mttr) and the invariant mass of 
the two jets from the hadronically decaying $W$ boson~($\mjj$).  We have a complete reconstruction of the \ttbar kinematics in the lepton+jets channel~\cite{tmt1,tmt2} with constraints from the precisely known $W$ boson mass and requiring the $t$ and $\bar{t}$ masses to be the same. 
Assuming that the leading four jets in the 
detector originate from the \ttbar decay products, there are twelve, six,  and two assignments 
of jets to quarks for 0-tag, 1-tag, and 2-tag events, respectively.
A minimization is performed for each assignment using a $\chi^2$ comparison to the \ttbar hypothesis with \mtr taken from the assignment that yields 
the lowest $\chi^2$~($\chi^{2}_{\mathrm{min}})$. To increase the statistical power of the measurement, we employ an additional
observable \mttr from the assignment that yields the second lowest $\chi^{2}$. 
To reject poorly reconstructed events, we require $\chi^{2}_{\mathrm{min}}<3$ and $\chi^{2}_{\mathrm{min}}<9$ for 0-tag and tagged~(both 1-tag and 2-tag) events, respectively.
The dijet mass \mjj is calculated as the invariant mass of two non $b$-tagged jets that provides 
the closest value to the known $W$ boson mass of \gevcc{80.39}~\cite{wmass}. 
We apply boundary conditions on \mtr and \mttr~($100 < \mtr, \mttr < \gevcc{350}$) and also
$\mjj$~($60 < \mjj < \gevcc{110}$, $50 < \mjj < \gevcc{120}$, and $50 < \mjj < \gevcc{125}$ 
for 0-tag, 1-tag, and 2-tag events, respectively). 
The estimated number of background events and the observed numbers of events 
after event selection, $\chi^2$, and boundary requirements are 
listed in Table~\ref{table_background_lj}.

\begin{table*}[ht]
\begin{center}
\caption[Expected number of background and signal events in the lepton+jets channel]{
Expected and observed numbers of signal and background events assuming \ttbar production 
cross-section $\sigma_{\ttbar}$ = \pb{7.45} and \mtop = \gevcc{172.5}.
}
\begin{ruledtabular}
\label{table_background_lj}
\begin{tabular}{lccccc}
              &0-tag  &1-tagL &  1-tagT  & 2-tagL  & 2-tagT \\
\hline
$W$+jets      &  703 $\pm$ 199  &170 $\pm$ 60 & 102 $\pm$ 37 & 11.6 $\pm$ 4.9 & 8.4 $\pm$ 3.5\\
$Z$+jets      & 52.3 $\pm$ 4.4 &8.9 $\pm$ 1.1& 5.9 $\pm$ 0.7&  0.8 $\pm$ 0.1 & 0.5 $\pm$ 0.1\\
Single top    &  4.8 $\pm$ 0.5  &10.5 $\pm$ 0.9&6.8 $\pm$ 0.6& 2.2 $\pm$ 0.3 & 1.7 $\pm$ 0.2 \\
Diboson       & 60.3 $\pm$ 5.6 &111 $\pm$ 1.4 &8.5 $\pm$ 1.1& 1.0 $\pm$ 0.2 & 0.8 $\pm$ 0.1\\
Multijets     & 143  $\pm$  114  &34.5 $\pm$ 12.6& 20.7 $\pm$ 16.6&4.4 $\pm$ 2.5&2.5 $\pm$ 2.4\\
\hline
Background    & 963 $\pm$ 229& 235 $\pm$ 61& 144 $\pm$ 41 & 19.9 $\pm$ 5.5 &13.8 $\pm$ 4.2\\
\ttbar signal & 645 $\pm$ 86 &695 $\pm$ 87&867 $\pm$ 108&192 $\pm$ 30&304 $\pm$ 47\\
\hline
\hline
Expected      &1608 $\pm$ 245&930 $\pm$ 106&1011 $\pm$ 115&212 $\pm$ 30&318 $\pm$ 47\\
Observed      &  1627     &    882   &    997     & 208      & 275 \\
\end{tabular}
\end{ruledtabular}
\end{center}
\end{table*}

We estimate the probability density functions~(p.d.f.s) of signal and
background using a kernel density
estimation method~\cite{KDEHEP}. A three-dimensional kernel density estimation~\cite{tmt4} accounts for the correlation between the three observables. 
The dijet mass \mjj of the two jets assigned to the $W$ in the lepton+jets channel is used for {\it in situ} calibration of the jet energy scale~(JES)~\cite{tmt1,tmt2}. 
The p.d.f.s for the observables are estimated at 76 discrete values of \mtop from \gevcc{130} to \gevcc{220} and at 29 discrete values of \djes from $-$3.0~$\sigma_{c}$ to 3.0~$\sigma_c$, where $\sigma_c$ is the CDF JES fractional uncertainty based on a combination of instrumental calibration and analysis of data control samples~\cite{jes}. The parameter \djes is defined as the difference between the true jet energy and our measured jet energy after calibration by a factor of $1+\djes\cdot\sigma_c$~\cite{tmt1,tmt2}. 
We interpolate the MC distributions 
to find p.d.f.s for arbitrary values of \mtop and \djes using the local polynomial smoothing
method~\cite{lps}.  
We fit the signal and background p.d.f.s to the distributions of the observables in the data using
an unbinned maximum likelihood fit~\cite{eml,minuit}. 
Independent likelihoods are used for each sub-sample, 0-tag, 1-tagL, 1-tagT, 2-tagL, and 2-tagT, and the total likelihood is obtained by multiplying them together~\cite{tmt2,tmt3,tmt4}. 
References~\cite{tmt2,tmt3,tmt4} provide detailed information about this technique.

We test the mass determination using 1~500 statistical trials
for a set of 11 different \mtop values ranging from \gevcc{160} to \gevcc{185}. 
In each experiment, we draw the number of signal and background events each from a
Poisson distribution centered at the expected number of signal and total background shown in Table~\ref{table_background_lj}, respectively. 
The distributions of the average deviation from the input top mass and the width of the deviation normalized to the estimated uncertainty for simulated experiments are corrected to be unity and zero, respectively. 
The correction is $M_{\mathrm{t}}^{\mathrm{corr}}=1.03\times M_{\mathrm{t}}^{\mathrm{meas}}-\gevcc{4.88}$, 
where $M_{\mathrm{t}}^{\mathrm{meas}}$ is the maximum likelihood estimate and $M_{\mathrm{t}}^{\mathrm{corr}}$ is the corrected value of the measurement. 
We increase the measured uncertainty by 2.9\% to correct the width of the pull. 
We also test the mass fit results using different values of \djes between $-$1.0~$\sigma_{c}$ to 1.0~$\sigma_c$ with three different \mtop points, 168, 173, and \gevcc{178}. With a correction discussed above, the residuals of \mtop from different \djes values are consistent with zero. 

\begin{table}
\begin{center}
\caption[Summary of systematics]{Estimated
systematic uncertainties (units in \gevccnoarg).}
\begin{ruledtabular}
\label{systtablesummary}
\begin{tabular}{lc}
Source              & Systematic uncertainty \\
\hline
Residual jet energy scale          & 0.52 \\
Signal modeling                    & 0.57 \\
$b$ jet energy scale               & 0.18 \\
$b$ tagging efficiency             & 0.03    \\
Initial and final state radiation  & 0.06 \\
Parton distribution functions      & 0.08 \\
Gluon fusion fraction              & 0.03 \\
Lepton energy scale                & 0.03 \\
Background shape                   & 0.20 \\
Multiple hadron interaction        & 0.07 \\
Color reconnection                 & 0.21 \\
MC statistics                      & 0.05 \\
\hline
Total systematic uncertainty       & 0.85 \\
\end{tabular}
\end{ruledtabular}
\end{center}
\end{table}

\begin{cfigure1c}
\begin{tabular}{ccc}
\includegraphics[width=0.32\textwidth]{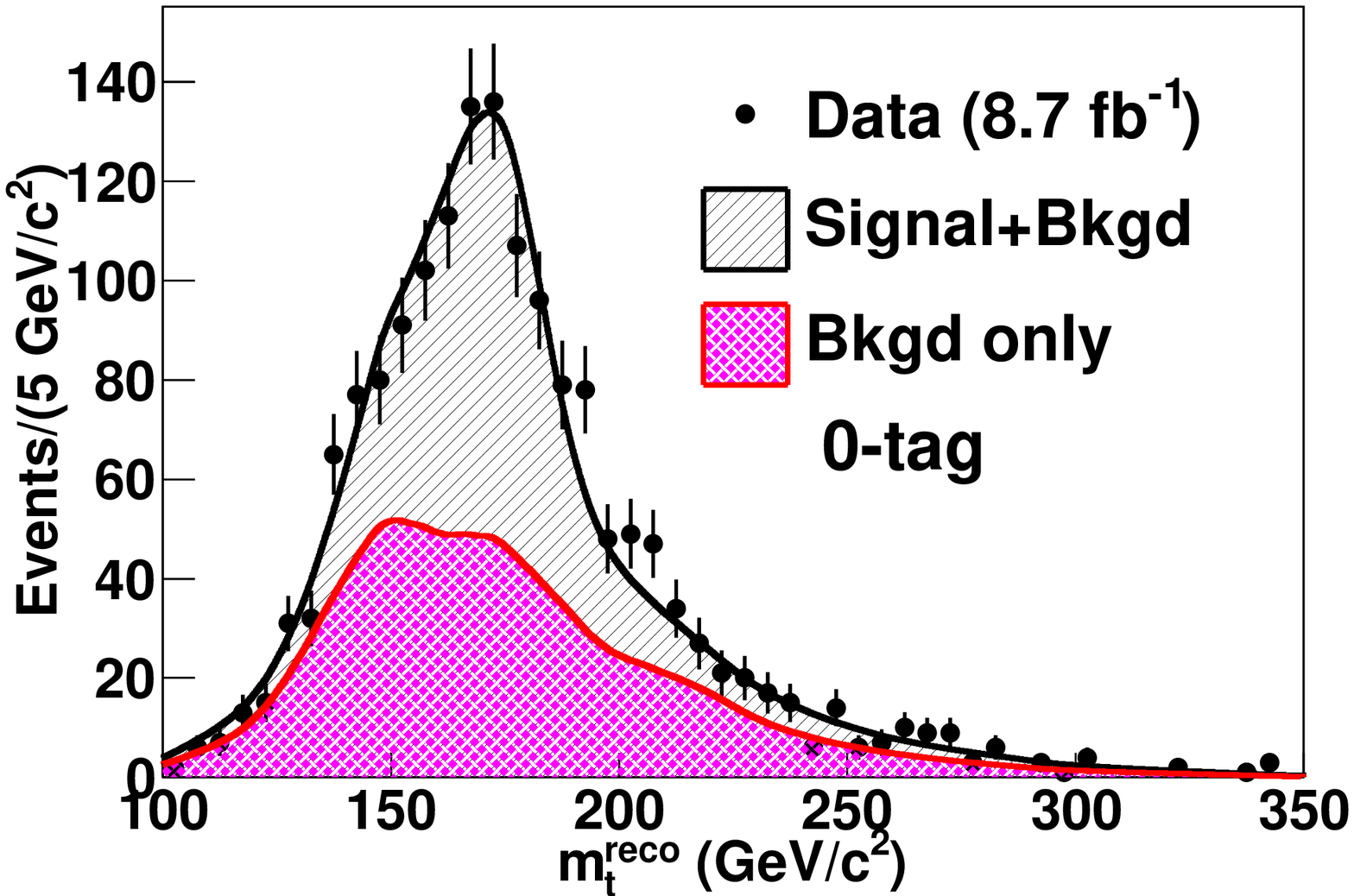}&
\includegraphics[width=0.32\textwidth]{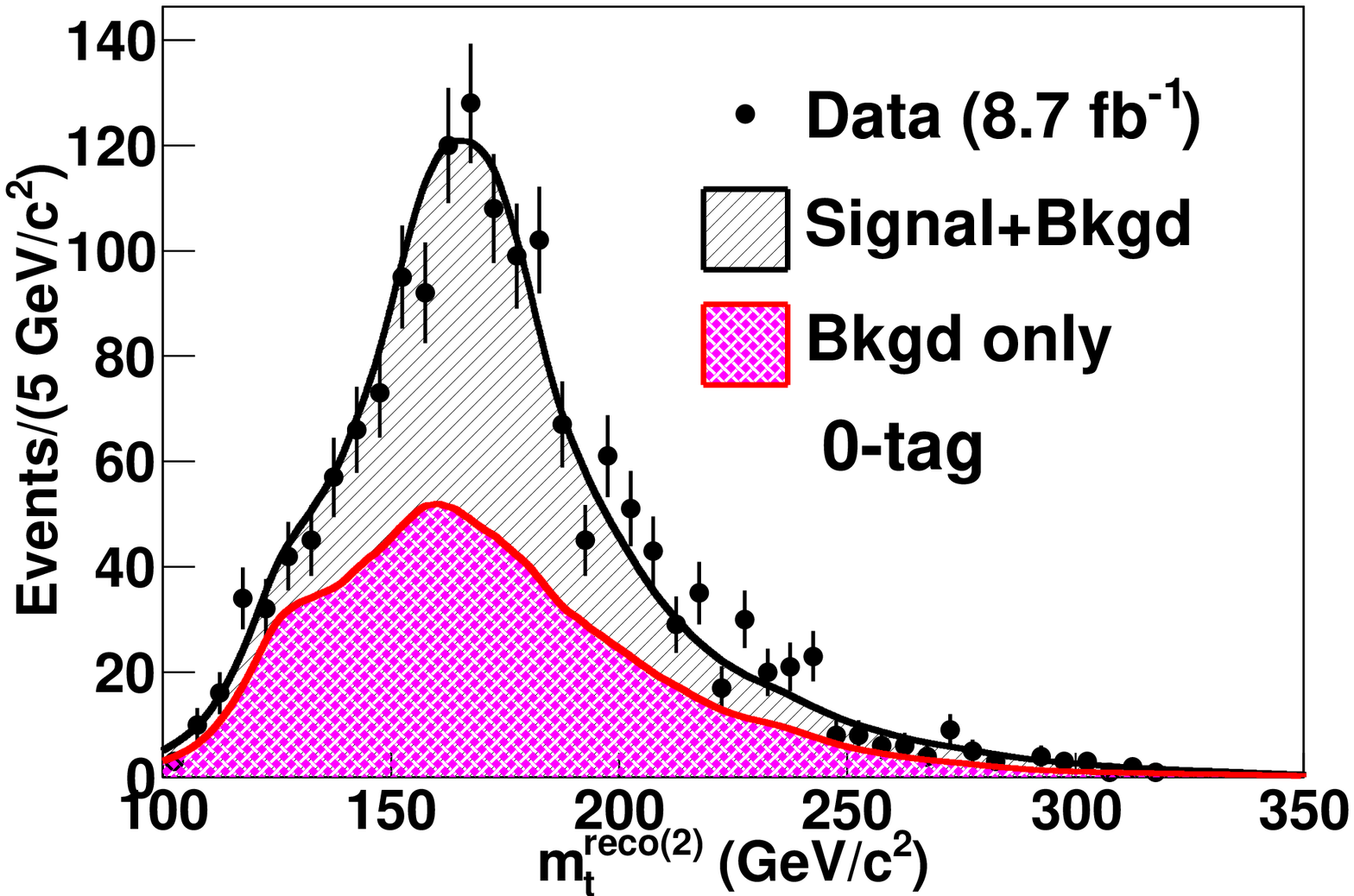}&
\includegraphics[width=0.32\textwidth]{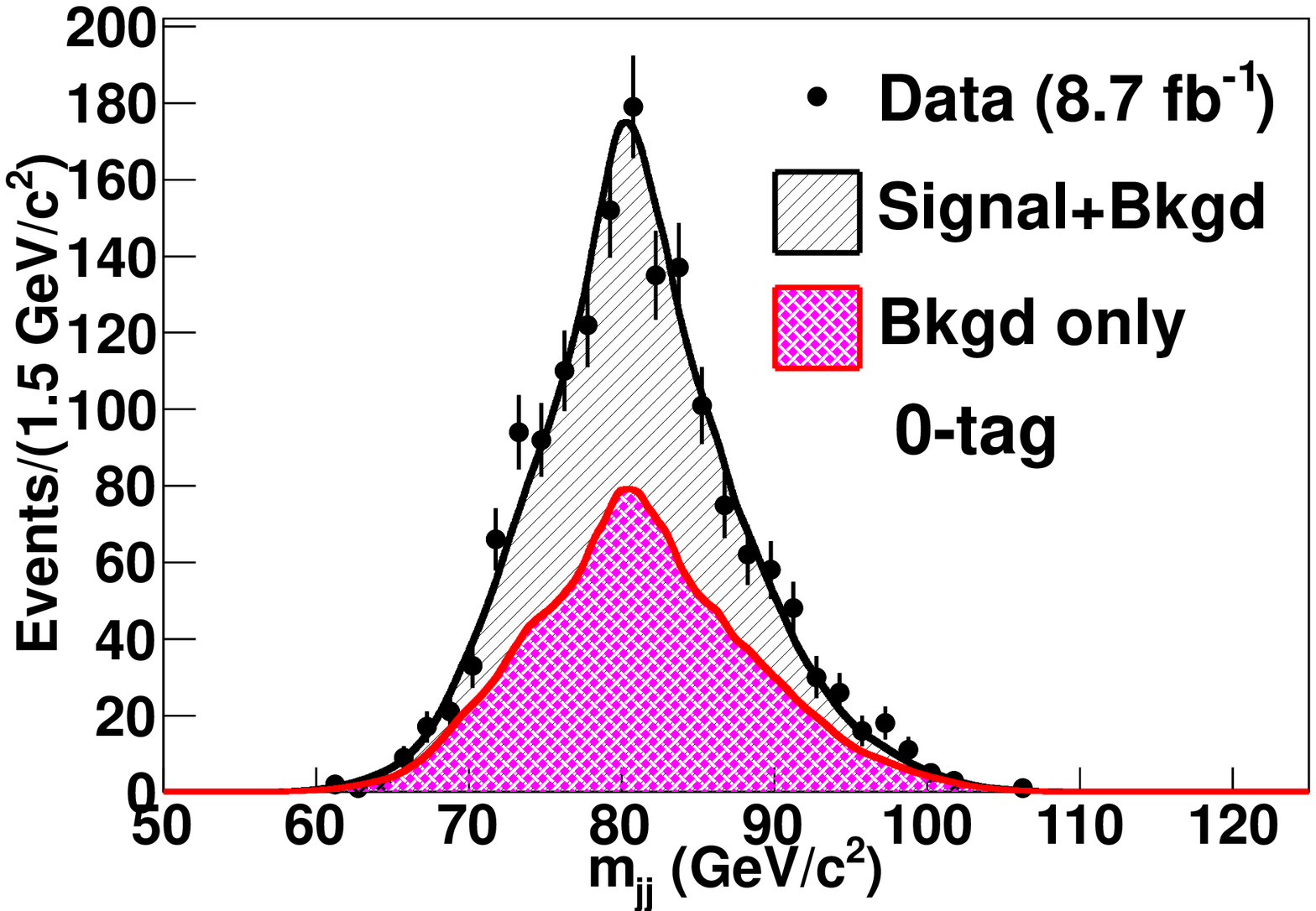}\\
\includegraphics[width=0.32\textwidth]{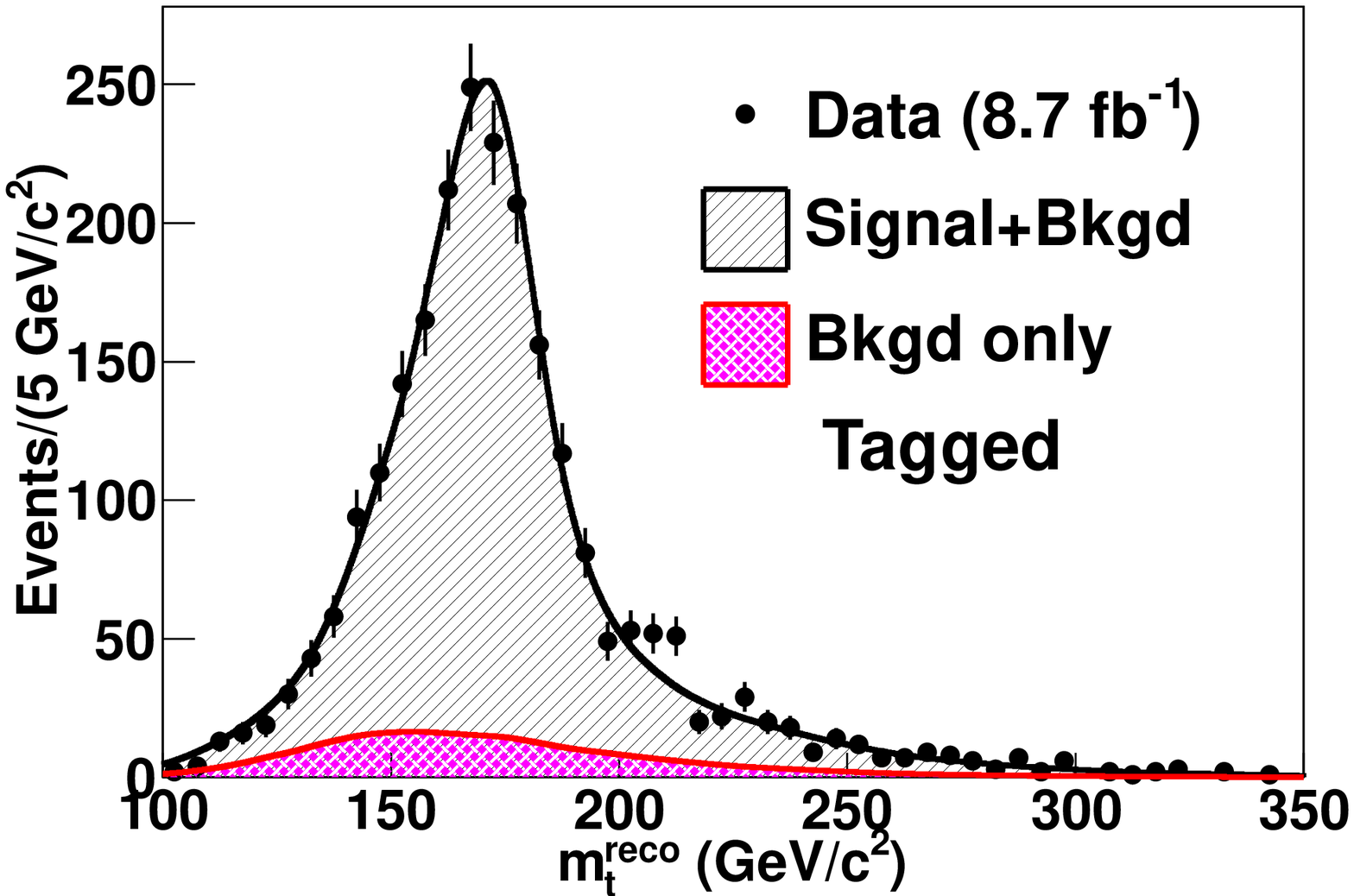}&
\includegraphics[width=0.32\textwidth]{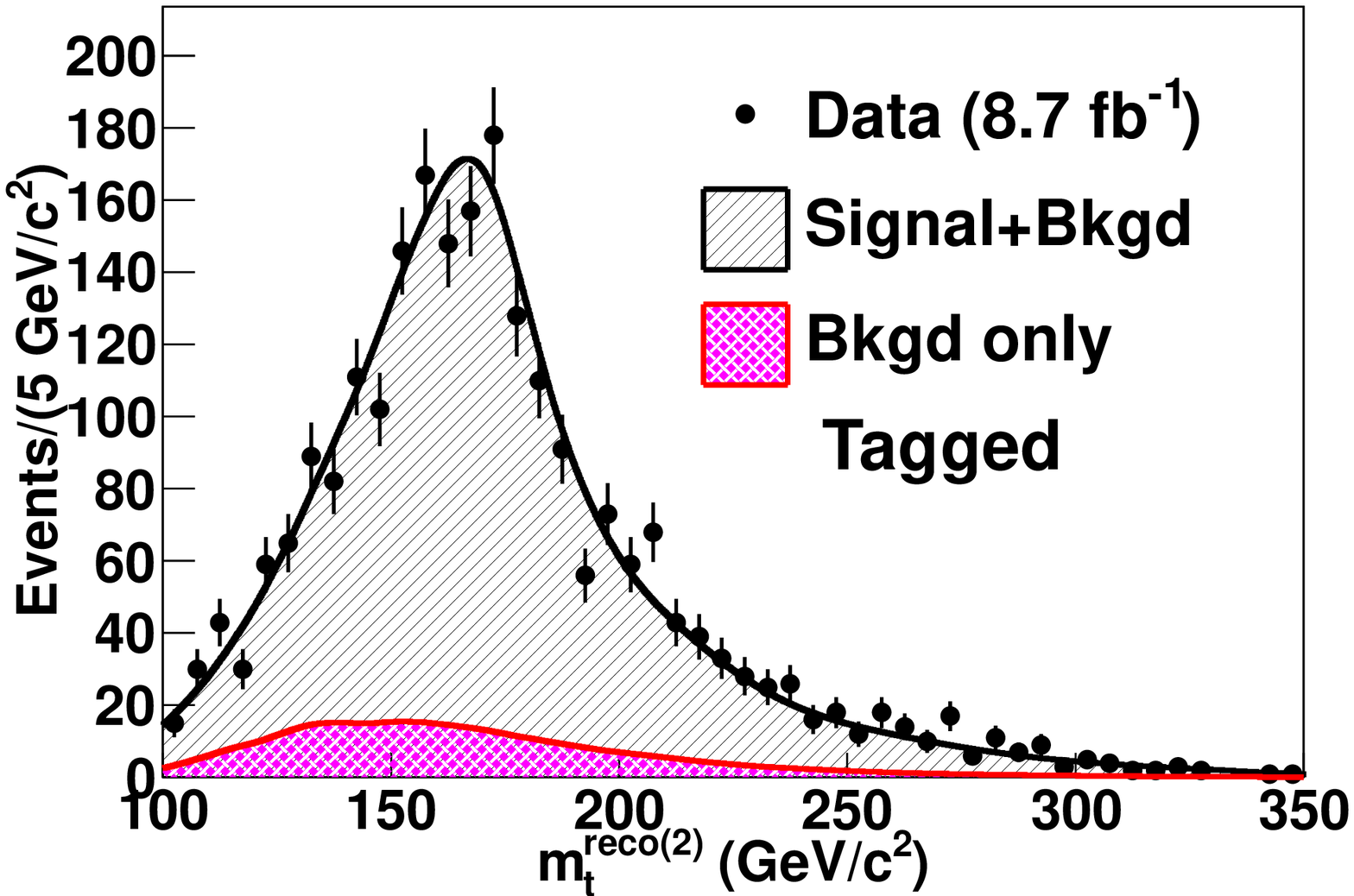}&
\includegraphics[width=0.32\textwidth]{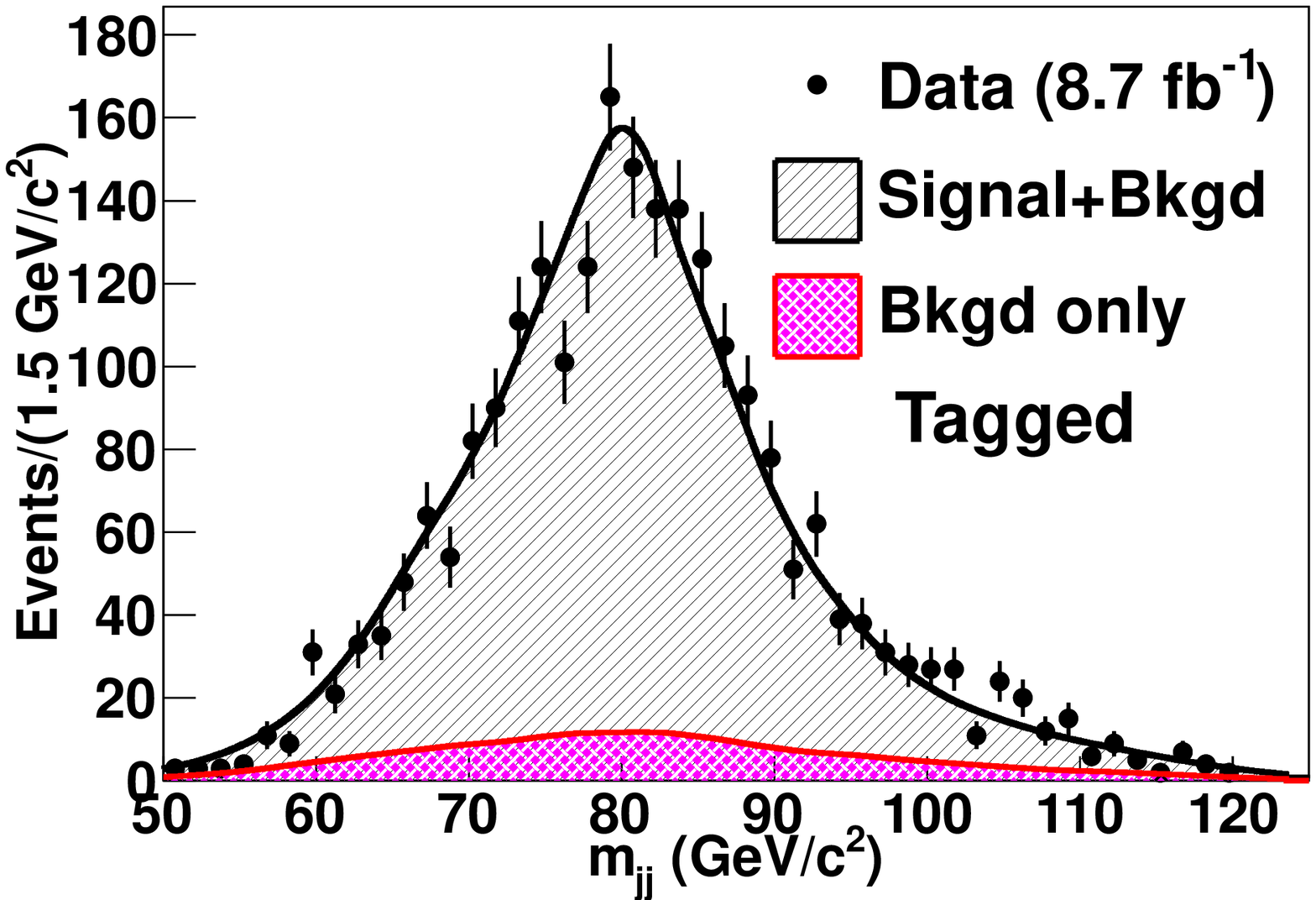}\\
\end{tabular}
\caption[Data fit]{Distributions of the three variables \mtr, \mjj, and \mttr, used to measure \mtop
	for 0-tag and tagged events. The data are overlaid with the 
predictions from the kernel density estimation probability distributions using \mtop = \gevcc{173} 
and the full background model. 
}
\label{fit_lj}
\end{cfigure1c}

We examine various sources of systematic uncertainties 
by comparing the analysis results of statistical trials in which we vary relevant parameters 
within their uncertainties.  The dominant sources 
are the residual JES~\cite{jes} and signal modeling. 
We vary the JES parameters within their uncertainties in both signal and background MC 
generated events and interpret the deviations of the results as additional uncertainties. 
The uncertainty arising from the choice of MC generator~(signal modeling) is estimated
by comparing the results of pseudoexperiments generated with 
{\sc pythia}~\cite{pythia} and {\sc herwig}~\cite{herwig}. We examine the effects of higher-order corrections using {\sc mc@nlo}~\cite{mcnlo}, a full next-to-leading-order simulation, which is added in the signal modeling. 
The systematic uncertainty to the energy corrections of $b$ jets~($b$-JES) arising from our modeling of $b$ fragmentation, 
$b$ hadron branching fractions, and calorimeter response, captures the additional 
uncertainty not taken into account in the residual JES.  
The uncertainty on the $b$-tagging efficiency can propagate into a bias of the \mtop measurement which is taken as a systematic uncertainty. 
The uncertainty due to the limited knowledge of initial-state radiation is constrained by studies of radiation in Drell-Yan events. We vary both initial and final state radiation within these constraints by extrapolating in the $p_{T}$ of Drell-Yan events to the \ttbar mass region~\cite{tmt1}.
We estimate the systematic uncertainty due to parton distribution functions
by varying the independent eigenvectors of the
 CTEQ6M~\cite{refcteq6m} parton distribution functions, varying the QCD scale~($\Lambda_{QCD}$), and comparing our nominal CTEQ5L~\cite{refcteq5l} with MRST72~\cite{refmrst} parton distribution functions. 
We vary the gluon fusion fraction from 5\% to 20\%, corresponding to the one standard deviation upper bound on the gluon fusion fraction~\cite{ggratio}. 
We estimate systematic uncertainties due to 
the lepton energy and momentum scales by propagating shifts in electron energy and muon 
momentum scales within their uncertainties. The background shape systematic uncertainty 
accounts for the variation of the background composition. We estimate a multiple-hadron-interaction systematic uncertainty to account for the difference in 
the average number of interactions between simulation and data. 
The color reconnection systematic uncertainty~\cite{CR} is evaluated using samples simulated with and without color reconnection effects with different {\sc pythia} tunes~\cite{CRtune}. All systematic uncertainties are summarized in Table~\ref{systtablesummary}.
The total systematic uncertainty adding individual components in quadrature 
is \gevcc{0.85}. The details of systematic uncertainty evaluations are in Ref.~\cite{tmt1,tmt2}.

We perform the likelihood fit to the data and apply the corrections obtained using the simulated experiments, and 
measure 
\begin{linenomath*}
\begin{eqnarray*}
\mtop & =& \gevcc{\measStatSyst{172.85}{0.71}{0.85}}.
\end{eqnarray*}
\end{linenomath*}
Figure~\ref{fit_lj} shows the distributions of the observables 
used for the \mtop measurement in the lepton+jets channel, overlaid with density estimates using \ttbar 
signal events with \mtop = \gevcc{173}~(close to the measured \mtop) and the full background model. 

In conclusion, we have performed a measurement of the top-quark mass using the template method in the lepton+jets using the full CDF Run~II data set corresponding to \invfb{8.7} \ppbar collisions. 
The result, $\mtop = \gevcc{\measErr{172.85}{1.11}}$, is the best single measurement of this important physics parameter. It is
consistent with the most recent Tevatron average of $\mtop = \gevcc{173.18 \pm 0.94}$~\cite{top_average} and will significantly contribute to the Tevatron's and world's average value for the top quark mass. 

\begin{acknowledgments}
	We thank the Fermilab staff and the technical staffs of the participating institutions for their vital contributions. This work was supported by the U.S. Department of Energy and National Science Foundation; the Italian Istituto Nazionale di Fisica Nucleare; the Ministry of Education, Culture, Sports, Science and Technology of Japan; the Natural Sciences and Engineering Research Council of Canada; the National Science Council of the Republic of China; the Swiss National Science Foundation; the A.P. Sloan Foundation; the Bundesministerium f\"ur Bildung und Forschung, Germany; the Korean World Class University Program, the National Research Foundation of Korea; the Science and Technology Facilities Council and the Royal Society, UK; the Russian Foundation for Basic Research; the Ministerio de Ciencia e Innovaci\'{o}n, and Programa Consolider-Ingenio 2010, Spain; the Slovak R\&D Agency; the Academy of Finland; and the Australian Research Council (ARC). 
\end{acknowledgments}


\begin{thebibliography}{99}
\bibitem{isolated} A lepton is isolated when $\left( \ptt - \ptl \right)/\ptl < 0.1$, where $\ptt$ is the total transverse momentum (energy) and $\ptl$ is the lepton transverse momentum (energy) for muon (electron) in a
  cone of radius $\Delta R \equiv \sqrt{(\Delta\eta)^{2}+(\Delta\phi)^{2}} = 0.4$~\cite{cdfco} with axis along the direction of the lepton.
\bibitem{cdfco} We use a right-handed spherical coordinate system with the origin at the center of the detector with the $z$-axis along the proton beam and the $y$-axis pointing up. $\theta$ and $\phi$ are the polar and azimuthal angles, respectively.  The pseudorapidity is defined by $\eta = -\ln \tan(\theta/2)$.  The transverse momentum and energy of a detected particle or jet are defined by 
$\pt = p \sin(\theta)$ and $\et = E \sin(\theta)$, respectively, where $p$ and $E$ are the momentum and energy of the particle.
\bibitem{met} The missing transverse energy, an imbalance of energy in the transverse plane of the detector, is defined by 
$\met = \vert \sum_{towers}\et{\hat n_{T}} \vert$, where ${\hat n_{T}}$ is the unit vector normal to the beam and pointing to a given calorimeter tower and $\et$ is the transverse energy measured in that tower.
\bibitem{Htcut} We require $\Ht > \gev{250}$ for 0-tag and 1-tag events. This requirement is not applied to the 2-tag events because of small background contribution in these sub-samples.
\end{thebibliography}
\end{document}